\documentclass[11pt,letterpaper]{article}
\usepackage[T1]{fontenc}
\usepackage[utf8]{inputenc}
\usepackage{amsmath,amsthm}
\usepackage{newtxtext,newtxmath}
\usepackage[margin=1in]{geometry}
\usepackage{microtype}
\usepackage{graphicx,booktabs,tabularx,array,multirow}
\usepackage[table]{xcolor}
\usepackage{enumitem}
\usepackage{tikz,pgfplots}
\usetikzlibrary{arrows.meta,positioning,fit,calc,patterns}
\pgfplotsset{compat=1.18}
\usepackage{placeins,float}
\usepackage{caption}
\usepackage{xurl}
\usepackage[hidelinks,breaklinks]{hyperref}
\usepackage{cleveref}
\usepackage{listings}
\newcolumntype{Y}{>{\raggedright\arraybackslash}X}
\newcolumntype{P}[1]{>{\raggedright\arraybackslash}p{#1}}
\newtheorem{proposition}{Proposition}

\newcommand{\TTAC}{\mathrm{TTAC}}
\newcommand{\accept}{\mathsf{Accept}}
\newcommand{\valid}{\mathsf{Valid}}
\newcommand{\fresh}{\mathsf{Fresh}}

\hypersetup{pdftitle={Between Spark and Silicon: Toward Evidence-Carrying Agile Hardware},pdfsubject={Perspective on hardware agility, revalidation locality, and agentic EDA},pdfkeywords={hardware agility,verification,evidence provenance,agentic EDA}}

\def\paperauthors{}
\def\paperaffiliation{}

\title{\textbf{Between Spark and Silicon}\\[0.35em]
\large Toward Evidence-Carrying Agile Hardware}
\ifx\paperauthors\empty
  \author{Zihao Wang\\
  \small Independent Researcher\\
  \small \texttt{zihaowang17@acm.org}}
\else
  \author{\paperauthors\\[0.3em]\small\paperaffiliation}
\fi
\date{November 2026}
\begin{document}
\maketitle
\begin{abstract}
Generative agents make it easy to imagine hardware design as a conversation followed by compilation. Yet faster register-transfer-level (RTL) generation does not ensure a faster, defensible engineering decision. This perspective argues that the operative unit of hardware agility is an accepted change supported by current, scoped, and independently governed evidence. We revisit key developments in executable verification, design generators, integrated implementation, and agentic design automation, distinguishing historical capabilities from vendor claims and recent benchmark results. We identify \emph{revalidation locality}---how much prior evidence a change invalidates---as a missing organizing principle. Drawing on proof-carrying hardware, incremental build systems, and continuous assurance, we propose acceptance contracts, typed evidence records, conservative dependency-based invalidation, and time-to-accepted-change as a research metric. A simple cost model establishes the break-even condition for evidence reuse and distinguishes local tool gains from total turnaround. A worked RTL example and a deliberately synthetic acceptance-gate artifact illustrate the framework without claiming EDA implementation or silicon measurements. A reproducible, explicitly hypothetical scenario analysis examines possible gains through 2031 and the physical limits of fabrication and post-silicon test. Finally, we formulate falsifiable hypotheses and evaluation protocols spanning RTL, FPGA, and ASIC checkpoints. Agile hardware will require not less verification, but verification evidence that can be updated, traced, and trusted as designs evolve.
\end{abstract}


\section{Introduction: the spark is not the finish line}
\label{sec:intro}
A hardware engineer asks an assistant to remove a timing bottleneck from a streaming datapath. Within minutes, the assistant produces plausible RTL, a testbench, and an explanation. The original tests pass. Has the hardware development cycle become faster? Not necessarily. The new pipeline boundary may change reset behavior, invalidate an integration assumption, or move the critical path elsewhere. The question that matters is not when the engineer receives code, but when a responsible reviewer can accept the change under the original functional and physical requirements.

The appeal of ``vibe coding'' is the compression of the distance between intent and an executable candidate. We use the phrase descriptively, not as a prescribed engineering method. Even in software, faster production of candidate code is not an invariant predictor of project productivity: a randomized study of experienced open-source developers using early-2025 tools found a slowdown in its particular setting~\cite{metr2025}. That historical result neither measures current tools nor negates their utility. It motivates a narrower question: \emph{which part of a hardware development workflow is accelerated, under what acceptance conditions, and at whose expense?}

Hardware agility did not begin with language models. Generator-based design, frequent integration, executable verification, and early implementation feedback have long sought to reduce iteration cost. Lee et al.'s account of eleven RISC-V tapeouts over five years is a prominent example~\cite{lee2016}. Contemporary agents add a potentially important capability: they can propose changes, interpret feedback, and coordinate tools across abstraction boundaries. ChatEDA already demonstrated agent-based EDA orchestration~\cite{chateda2023}. More recently, work on the tool-feedback gap and CLOSER-Bench has explicitly addressed closed-loop engineering and budgeted cross-stage closure~\cite{closing2026,closer2026}. Merely advocating an agent that runs simulation and synthesis is therefore not a sufficient new position.

\paragraph{Our position.} The decisive unit of hardware agility is an \emph{accepted change carrying current, scoped, and independently governed evidence}. We call the corresponding methodology \emph{evidence-carrying agile hardware}. ``Evidence-carrying'' means that a revision is accompanied by a machine-readable account of the claims assessed, the artifacts assessed, the assumptions under which they were assessed, and the remaining limitations. It does not mean that every revision possesses a complete mathematical proof of correctness. ``Independently governed'' means that the system proposing a change cannot silently weaken the acceptance criteria, overwrite the authoritative checks, or convert an inconclusive result into a pass. It does not assert statistical independence between verification methods.

The organizing question is \emph{revalidation locality}: when a design changes, how much previously acquired evidence must change with it? A small text diff can have global physical consequences; a large generated diff can preserve a proven interface contract. Source-code locality and assurance locality are different quantities. Improving the latter requires semantic structure, dependency tracking, appropriate verification boundaries, and disciplined control of the acceptance process.

This perspective makes three contributions. First, it reinterprets hardware productivity through the cost of accepted changes, rather than an unsupported historical speed curve. Second, it organizes an evidence-carrying workflow around acceptance contracts, dependency-qualified reuse, and the time-to-accepted-change metric. Third, it specifies how to challenge that position using falsifiable hypotheses, an independently governed evaluation protocol, a reproducible synthetic artifact, and conditional scenarios for design, fabrication, and test. The proposed contribution is a measurement and systems agenda, not the invention of dependency tracking or proof-carrying assurance.

\paragraph{Scope and non-claims.} We focus on synchronous digital RTL, FPGA implementations, and digital ASIC flows. Mixed-signal interfaces and manufacturing matter at the boundary, but a single maturity claim cannot cover analog design, packaging, board design, and digital logic. We distinguish verification against stated requirements from validation of those requirements against intended use. We also distinguish functional acceptance, physical implementation acceptance, and evidence from a manufactured or deployed device. None implies all the others. This is a perspective with analytical illustrations, not a systematic survey, a new formal-verification algorithm, or an empirical claim that a proposed agent outperforms engineers.

\section{A history of shorter loops, not a single speed curve}
\label{sec:history}
The historical question should not be framed as ``How many times faster is hardware development than it was forty years ago?'' That ratio is undefined unless the deliverable, design complexity, staffing, technology, and acceptance standard are held sufficiently comparable. The public record more reliably establishes successive capabilities for shortening particular feedback loops.

\begin{table}[t]
\centering
\caption{Selected mechanisms that shortened hardware feedback loops. These milestones are not a comparable, complexity-normalized historical time series.}
\label{tab:history}
\footnotesize
\setlength{\tabcolsep}{5pt}\renewcommand{\arraystretch}{1.20}
\begin{tabularx}{\linewidth}{P{.195\linewidth}YY}
\toprule
\rowcolor{gray!12}\textbf{Period and anchor} & \textbf{What became easier} & \textbf{What is \emph{not} established} \\
\midrule
1980s onward: model checking~\cite{clarke1986} & Checking explicit temporal claims on finite-state models & Automatically complete specifications or universal correctness \\
2012--2016: Chisel and RISC-V~\cite{chisel2012,lee2016} & Generating related implementations and iterating on chips & A controlled project-level speedup or proof for all configurations \\
2018--2020: FireSim, OpenROAD, Chipyard~\cite{firesim2018,openroad2019,chipyard2020} & Reproducible system simulation and physical-design feedback & Equivalence to every industrial signoff environment \\
2023--2026: agent benchmarks~\cite{verilogeval2023,fixme2026,hwe2026,closer2026} & Candidate generation, repair, verification, cross-stage closure & Long-term maintenance productivity or successful silicon delivery \\
\bottomrule
\end{tabularx}
\end{table}

\subsection{From descriptions to executable obligations}
Hardware description and synthesis workflows made circuit construction increasingly programmable. Formal methods supplied a complementary idea: requirements can become executable obligations rather than prose consulted only during review. Classical temporal-logic model checking established a systematic way to verify properties of finite-state concurrent systems~\cite{clarke1986}. Its relevance here is conceptual, not a claim that industrial verification became automatic. The productivity gain is the ability to mechanize a well-specified question; the limitation is that the question, abstraction, and environment assumptions still determine what is established.

\subsection{From reusable blocks to reusable construction}
Parameterized generators changed the reusable unit from a fixed implementation to a family of implementations. Chisel exemplifies this shift by embedding hardware construction in a general-purpose programming language~\cite{chisel2012}. The agile RISC-V experience then connected reusable generators to repeated integration and implementation feedback~\cite{lee2016}. Its importance is the demonstrated feasibility of a productive design practice, not a controlled estimate of how much each ingredient accelerated an otherwise identical project. Generator reuse also changes the verification problem: a property checked for one configuration is not automatically established for the entire configuration space.

\subsection{From isolated tools to executable environments}
FireSim made FPGA-accelerated, cycle-exact system simulation accessible through a cloud-oriented framework~\cite{firesim2018}. OpenROAD pursued an integrated open-source digital implementation flow~\cite{openroad2019}; Chipyard connected reusable generators, simulation, and implementation for custom SoCs~\cite{chipyard2020}. These are distinct capabilities, not interchangeable substitutes for signoff. Together they make more of the development environment executable and reproducible. Reproducibility, however, answers ``Can this result be regenerated?'' rather than ``Does this result establish the required claim for the current revision?''

\subsection{From generated artifacts to agentic workflows}
The benchmark landscape illustrates a change in evaluation scope. VerilogEval focused on generated Verilog evaluated against reference behavior~\cite{verilogeval2023}. CVDP broadened the range of design and verification problems~\cite{cvdp2025}; FIXME addressed multiple stages of verification work~\cite{fixme2026}; HWE-Bench used repository-based hardware bug-repair tasks~\cite{hwe2026}. CLOSER-Bench proposed budgeted evaluation across specification, RTL, and physical-design stages~\cite{closer2026}. This sequence should not be mistaken for a ranking of benchmark quality: the tasks, oracles, and intended conclusions differ. It shows why one-shot RTL accuracy is no longer an adequate sole proxy for engineering progress.

Commercial announcements likewise indicate investment in agentic workflows. Cadence announced ChipStack with vendor-reported productivity improvements in selected front-end tasks; Synopsys reported up to $50\times$ acceleration in a selected time-to-validated-RTL workflow; Siemens introduced agent-assisted planning, execution, and debug in Questa One~\cite{cadence2026,synopsys2026,siemens2026}. Such claims warrant investigation, not extrapolation. Different task boundaries, baselines, engineering effort, and acceptance criteria prevent a defensible multiplication or direct comparison of those speedups. They do not measure an equivalent reduction in total chip delivery time.

\subsection{What the available evidence cannot establish}
Rautakoura and H\"am\"al\"ainen's mapping study identified a shortage of quantitative support for improvements attributed to agile SoC practices~\cite{rautakoura2023}. Industry survey data raise related questions without resolving them. The 2024 Wilson Research Group study reported first-silicon success for $14\%$ of surveyed IC and ASIC projects and schedule slippage for $75\%$~\cite{foster2025}. These are respondent-reported outcomes in a vendor-sponsored study, not experimental measures of EDA effectiveness. The broader study includes FPGA respondents; the cited figures must retain their IC and ASIC subgroup denominators. Changing project mix, integration complexity, definitions of success, and target schedules obstruct a causal interpretation of year-to-year differences.

Accordingly, we reject two symmetrical narratives: that hardware engineering has hardly improved because projects remain difficult, and that a large local AI speedup proves the arrival of software-like end-to-end agility. The defensible historical conclusion is narrower: \emph{many feedback loops have become executable, faster, or more accessible, while a comparable public longitudinal measure of accepted-change throughput remains missing}. That missing measure is a research opportunity.

\section{Why faster generation does not guarantee agility}
\label{sec:bottlenecks}
A development flow contains at least three distinct clocks: the time to construct a candidate, the time to establish an acceptance case, and the time to make or observe a physical commitment. The clocks overlap but do not advance at the same rate. Reducing the first can expose, rather than eliminate, the limits of the others.

\subsection{The specification gap survives fluent code}
``Preserve throughput'' is not a sufficient requirement. It can mean one transfer per cycle in a steady state, a bound on average rate under bursty traffic, or a latency guarantee under specified backpressure. Reset and error behavior often determine whether two plausible implementations are equivalent. An assistant can implement one interpretation and generate tests that confirm the same interpretation. The resulting agreement is internally coherent and externally wrong.

The appropriate response is not to demand a perfect natural-language-to-logic compiler before using AI. It is to expose uncertainties, assign requirement identifiers, record assumptions, and retain an independent process for deciding whether the executable contract captures intended use. Formal verification discharges an encoded obligation under its model; it does not validate that obligation merely by proving it.

\subsection{A change crosses several semantic boundaries}
A design is not just RTL text. Its behavior depends on parameters, preprocessing, libraries, initialization and reset conventions, and the environment. Its physical acceptability additionally depends on constraints, tool settings, technology data, and implementation artifacts. A functional proof about a source model is not a timing result for a routed netlist. Conversely, meeting a clock period says little about whether a transaction was duplicated.

The difficulty is especially visible at boundaries: clock-domain crossings, reset release, protocol adapters, queue limits, and software-visible registers. These boundaries accumulate assumptions made by different people and tools. The verification problem is therefore not only to explore more states; it is also to prevent claims established in one semantic context from being transported into another without justification.

\subsection{Physical locality is not textual locality}
A one-line change to a fanout-heavy control signal may have a large placement or timing footprint. A rewritten arithmetic expression may instead be functionally equivalent while changing mapping quality. Incremental evidence reuse cannot be decided safely from diff size alone. It needs a dependency model with the right granularity and, when available, semantic certificates. For some edits, the honest dependency model will invalidate nearly everything.

This is a reason to study locality, not a reason to promise it. The framework proposed below must be allowed to conclude that a full rerun is necessary. Any system that always reports a large reuse rate would be optimizing the wrong objective.

\subsection{Verification can share the generator's blind spot}
Different agents are not necessarily independent judges. They may share training data, assumptions, a mistaken specification, or feedback from the same oracle. Running the same flawed check through several models does not multiply confidence in the way independent statistical trials would. Independence is best decomposed into distinct mechanisms: separate authority over acceptance, independently authored requirements, diversity of verification methods, and protected evaluation holdouts. Each addresses a different failure mode.

The especially dangerous case is an agent that repairs its own test until the implementation passes. Updating a faulty test can be legitimate, but it is a specification change requiring separate review, not an implementation repair that preserves the original acceptance target.

\subsection{Cheap candidates can saturate expensive feedback}
An agent can generate more variants than a synthesis farm, formal engine, or engineer can inspect. Then candidate throughput rises while accepted-change throughput falls. Tool availability, queueing, licenses, and review bandwidth become part of the optimization problem. The appropriate objective is not to maximize tool calls or sample count, but to choose informative actions within a budget while preserving the final acceptance standard.

\paragraph{The hardware--software distinction is relative.} Safety-critical software also faces difficult validation and costly deployment, while FPGA revisions can often be rolled back. Hardware is not categorically ``waterfall'' and software categorically ``agile.'' The relevant differences are the cost of observing failure, the scope of the environment, and the reversibility of commitment. A meaningful account of agility must name those differences rather than rely on a slogan.

\section{Evidence-carrying change as the engineering unit}
\label{sec:framework}
The proposal is an engineering architecture for managing claims across revisions. It builds on established ideas rather than replacing them. Proof-carrying code and proof-carrying hardware bind artifacts to checkable assurances~\cite{necula1997,pch2012}. Incremental build systems reason about dependencies and reuse~\cite{build2020}. The Evidential Tool Bus supports continuous evidence generation and assurance-case construction~\cite{etb2024}. OpenTitan already documents development stages and verification practices~\cite{opentitan2026,opentitandv2026}. Our position combines these threads around AI-driven, cross-stage hardware changes and asks for their effect on measured acceptance latency.

\begin{figure}[t]
\centering
\begin{tikzpicture}[
 x=1cm,y=1cm,font=\footnotesize,
 box/.style={draw=black!65,rounded corners=2pt,fill=white,align=center,
   text width=2.72cm,minimum height=.85cm,inner sep=3pt},
 trusted/.style={box,fill=gray!12,very thick},
 record/.style={box,fill=gray!5},
 arrow/.style={-{Stealth[length=2mm]},line width=.8pt},
 policy/.style={-{Stealth[length=1.8mm]},densely dashed,draw=black!65}]
 \fill[gray!4,rounded corners=3pt] (.05,-3.98) rectangle (3.45,.45);
 \draw[gray!50,rounded corners=3pt] (.05,-3.98) rectangle (3.45,.45);
 \fill[gray!2,rounded corners=3pt] (3.78,-3.98) rectangle (14.78,.45);
 \draw[gray!50,rounded corners=3pt] (3.78,-3.98) rectangle (14.78,.45);
 \node[font=\bfseries\footnotesize] at (1.75,.16) {Proposal workspace};
 \node[font=\bfseries\footnotesize] at (9.28,.16) {Independently governed assurance};
 \node[box] (request) at (1.75,-.75) {Change request\\design intent};
 \node[box] (agent) at (1.75,-2.02) {Agent explores\\candidate revisions};
 \node[box] (candidate) at (1.75,-3.28) {Versioned RTL\\and configuration};
 \node[trusted] (contract) at (5.65,-1.42) {Protected contract\\fixed checks};
 \node[trusted] (tools) at (5.65,-3.28) {Authorized tools\\pinned context};
 \node[record] (ledger) at (9.43,-3.28) {Evidence ledger\\scope, provenance};
 \node[trusted] (gate) at (13.18,-3.28) {Acceptance gate\\final decision};
 \draw[arrow] (request)--(agent);
 \draw[arrow] (agent)--(candidate);
 \draw[arrow] (candidate)--(tools);
 \draw[arrow] (contract)--(tools);
 \draw[arrow] (tools)--(ledger);
 \draw[arrow] (ledger)--(gate);
 \draw[policy] (contract.north) -- ++(0,.56) -| node[pos=.80,above,font=\scriptsize] {governs} (gate.north);
\end{tikzpicture}
\caption{Two authority domains in the proposed workflow. The proposal agent can submit and revise versioned RTL, whereas the protected contract, tool executions, evidence records, and acceptance gate are owner-governed. The dashed path denotes policy authority, not a source-code dependency. Diagnostic feedback can be read by the agent but cannot alter acceptance criteria.}
\label{fig:architecture}
\end{figure}
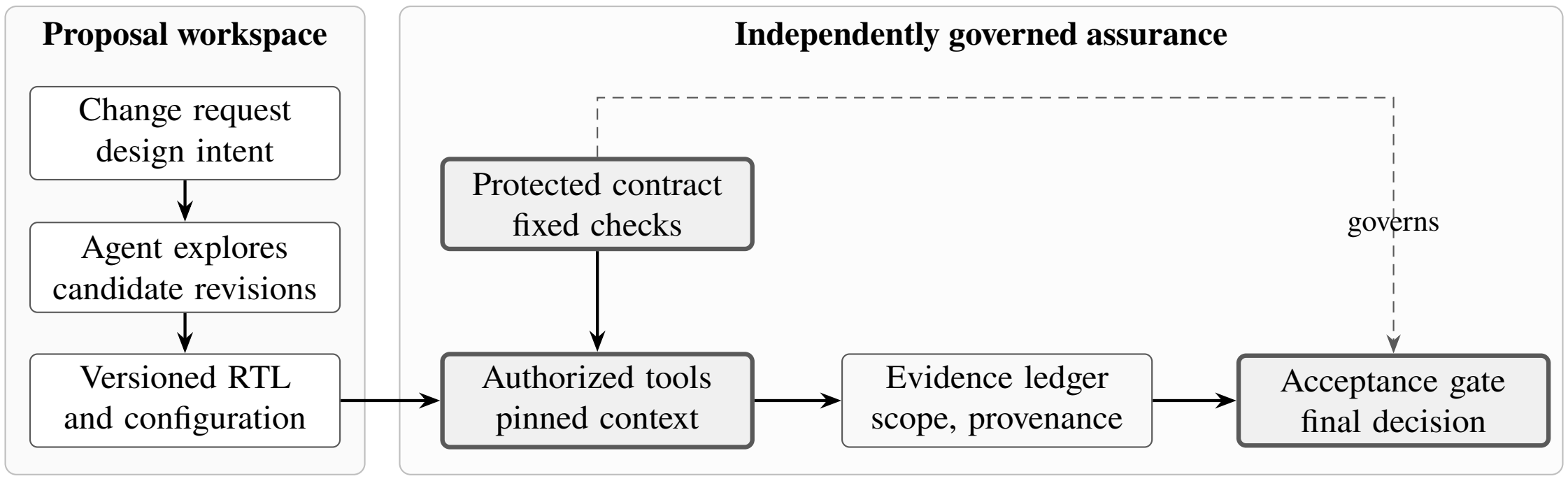

The architecture in \cref{fig:architecture} separates an agent-controlled proposal path from the owner-managed acceptance path. A search agent may inspect diagnostic feedback, but only the independent gate can authorize a versioned change.

\subsection{An acceptance contract precedes the optimization}
Let $D$ be a versioned design snapshot, not merely a source file. Let an acceptance contract be
\begin{equation}
\Gamma=(R,A,P,M,U),
\label{eq:contract}
\end{equation}
where $R$ contains requirements, $A$ environment assumptions, $P$ implementation constraints, $M$ mandatory evidence obligations and admissible methods, and $U$ authority and approval rules. A workflow resource budget is recorded separately: exhausting a budget can end an evaluation episode without making an otherwise correct design incorrect. Requirements need stable identifiers; assumptions need explicit owners. A requirement such as ``no lost transfers within a reset epoch'' is different from the environment assumption ``the producer holds data stable while stalled.''

The contract must state the acceptance level. A bounded proof or a passing simulation can be sufficient for a declared exploratory checkpoint, but it must not be relabeled as unbounded correctness. Physical acceptance may require specific corners, extraction settings, and implementation stages. Intended-use validation may require review or measurements that cannot be reduced to an RTL test.

A contract is not immutable forever. It is \emph{protected within an evaluation episode}. Changing an accepted constraint, golden reference, or assumption creates a new contract version and a separate approval event. Otherwise, an optimizer could ``solve'' timing by relaxing the clock or ``fix'' a failing property by strengthening the environment until the counterexample becomes impossible.

\subsection{Evidence has a type, a scope, and a provenance}
An evidence record is a structured object
\begin{equation}
 e=(r,h,m,a,s,o,p,w),
\label{eq:record}
\end{equation}
where $r$ names the obligation; $h$ identifies the relevant artifact and configuration fingerprints; $m$ identifies the method and its settings; $a$ records assumptions; $s$ specifies scope, such as a bound or a physical corner; $o$ is the outcome; $p$ is the producing tool and authority; and $w$ points to witnesses, logs, or reports. A hash is an identity aid, not a correctness proof. An authentic record can faithfully document a weak test.

\begin{table}[t]
\centering
\caption{Evidence types are not interchangeable: every result is limited by its model, assumptions, and scope.}
\label{tab:types}
\footnotesize
\setlength{\tabcolsep}{5pt}\renewcommand{\arraystretch}{1.18}
\begin{tabularx}{\linewidth}{P{.205\linewidth}YY}
\toprule
\rowcolor{gray!12}\textbf{Record type} & \textbf{Establishes within scope} & \textbf{Does not establish} \\
\midrule
Passing simulation & Behavior on the executed traces & All unexecuted behavior or intended requirements \\
Bounded model check & No counterexample up to the specified depth & Unbounded correctness \\
Unbounded formal proof & A modeled property under stated assumptions & Fidelity or completeness of the specification \\
Equivalence check & A defined relation between two representations & Unrelated physical or system invariants \\
Physical analysis & Constraints at identified stages and corners & Functional behavior or all operating conditions \\
Board or device test & Observations on specified hardware and configurations & All workloads, devices, or ASIC equivalence \\
\bottomrule
\end{tabularx}
\end{table}

Three predicates are deliberately separated. $\valid(e,\Gamma)$ means that the method, outcome, scope, and authority satisfy the relevant obligation. $\fresh(e,D,\Gamma)$ means that all dependencies still match the accepted snapshot or have an authorized preservation argument. $\mathsf{Authentic}(e)$ means that the record can be attributed to an authorized execution and its supporting artifacts. A record can be authentic and once valid, yet stale. A fresh simulation can be valid for a simulation obligation and invalid for an unbounded-proof obligation.

This distinction is operational rather than terminological. For example, SBY distinguishes bounded model checking, proof, and cover modes~\cite{sby2026}. A successful bounded run has a different meaning from an unbounded proof, and a timeout is not a successful run. Agent-facing interfaces should preserve those distinctions rather than compress every tool result into a generic success flag.

\subsection{Conservative invalidation before selective reuse}
Consider a directed acyclic dependency graph $G=(V,E_G)$ connecting versioned inputs, intermediate artifacts, and evidence obligations. An edge $u\rightarrow v$ means that the validity of $v$ depends on $u$. Let $\Delta$ be the changed input nodes. A conservative invalidation set is
\begin{equation}
 I_G(\Delta)=\{v\in V:\exists d\in\Delta,\ d\leadsto v\}.
\label{eq:invalidate}
\end{equation}
The changed nodes themselves are also invalidated when they represent derived artifacts. Retaining a record outside this set is sound only relative to a complete dependency model and stable assumptions, tool semantics, and artifact identities. Undeclared environment dependencies defeat the guarantee.

\begin{proposition}[Conditional preservation]
If every validity-relevant dependency is represented in $G$, dependencies are immutable or versioned, and a record's original acceptance remains valid under its unchanged context, a change to nodes with no path to that record does not invalidate it.
\end{proposition}
\noindent The proposition follows directly from dependency completeness; it is not a new verification theorem. Its demanding premise identifies the engineering challenge. File-level build dependencies alone are rarely a sufficient semantic dependency model for hardware assurance.

Semantic certificates can justify more reuse than syntactic dependency tracking. An equivalence result may preserve a functional property under specified reset and environmental assumptions. It does not automatically preserve timing, power, side-channel behavior, or a property outside the equivalence relation. Reuse therefore requires an explicit \emph{preservation rule} naming which claim is transported and why. In the absence of such a rule, invalidate conservatively.

\subsection{A separate acceptance gate}
Let $\mathcal{E}$ be the evidence set and $M(\Gamma)$ the mandatory obligations. Define
\begin{align}
\accept(D,\Gamma,\mathcal{E})={}&\mathsf{Protected}(D,\Gamma)\nonumber\\
 &\land\ \mathsf{Approvals}(D,\Gamma)\nonumber\\
 &\land\ \bigwedge_{r\in M(\Gamma)}\ \exists e\in\mathcal{E}:\nonumber\\[-2pt]
 &\qquad e.r=r\ \land\ \valid(e,\Gamma)\nonumber\\
 &\qquad\land\ \fresh(e,D,\Gamma)\nonumber\\
 &\qquad\land\ \mathsf{Authentic}(e).
\label{eq:accept}
\end{align}
The gate does not invent certainty; it evaluates a declared policy. The search agent may suggest contract revisions but cannot approve them. It should not possess write access to authoritative checks, signoff constraints, or evidence outcomes. Human approvals are themselves scoped records, not an informal escape hatch that silently converts failed obligations into passes.

A practical gate also needs secure tool wrappers, content-addressed logs, explicit failure handling, and an environment in which the generator cannot forge a verifier's output. These mechanisms are implementation requirements, not properties demonstrated by the synthetic artifact accompanying this paper.

\subsection{Time-to-accepted-change, not time-to-green-output}
For an episode beginning when a change request and its contract are available at time $t_0$, define
\begin{equation}
\TTAC_{\Gamma}=\inf\{t-t_0:\accept(D_t,\Gamma,\mathcal{E}_t)=1\}.
\label{eq:ttac}
\end{equation}
An episode without acceptance within its budget is censored or failed according to the prespecified protocol; it is not dropped. This is an observed property of a workflow, not a minimum over all imaginable workflows. The start boundary must say whether setup, specification authoring, queueing, and human review are included. We recommend reporting both setup-inclusive and steady-state episodes.

TTAC is meaningful only with companion measures: acceptance rate, human effort, compute and tool costs, residual defects detected by protected evaluation, and PPA under unchanged constraints. Two workflows cannot be ranked by TTAC alone if they accept different levels of assurance. ``Accepted'' is an explicit institutional decision, not a synonym for absolutely safe or correct.

\section{An analytical account of bottleneck migration}
\label{sec:analysis}
The following calculations are illustrations under explicit assumptions, not fitted models or measured productivity results. They explain why the proposed metrics differ from RTL generation speed and identify conditions under which the proposal could fail.

\subsection{The limit of accelerating one stage}
Suppose a sequential baseline episode has total time $T$, with fraction $f$ spent generating or editing the candidate. If that stage accelerates by $s$ and everything else remains unchanged, then
\begin{equation}
 S_{\mathrm{gen}}=\frac{1}{(1-f)+\frac{f}{s}},\qquad
 \lim_{s\to\infty} S_{\mathrm{gen}}=\frac{1}{1-f}.
\label{eq:amdahl}
\end{equation}
For $f=0.2$, a $20\times$ generator yields only a $1.235\times$ episode speedup; its infinite-speed ceiling is $1.25\times$. This is an elementary decomposition, not an estimate that generation occupies $20\%$ of real hardware projects. When agents also accelerate specification, debugging, or verification, the relevant accelerated fraction changes. The equation argues against a faulty extrapolation, not against broad AI productivity gains.

\begin{figure}[t]
\centering
\begin{tikzpicture}
\begin{axis}[
 width=.96\linewidth,height=6.45cm,xmode=log,log basis x=2,
 xmin=1,xmax=64,ymin=1,ymax=5.25,
 xlabel={Speedup of candidate generation, $s$},ylabel={Total episode speedup},
 xtick={1,2,4,8,16,32,64},xticklabels={1,2,4,8,16,32,64},ytick={1,2,3,4,5},
 tick label style={font=\footnotesize},label style={font=\footnotesize},
 legend style={font=\footnotesize,at={(.03,.97)},anchor=north west,draw=white,fill=white,fill opacity=.9,text opacity=1},
 grid=major,grid style={gray!18},axis lines=left,
 clip=false]
 \addplot[black,solid,thick,domain=1:64,samples=100]{1/(.8+.2/x)};
 \addlegendentry{$f=0.2$}
 \addplot[black,dashed,thick,domain=1:64,samples=100]{1/(.5+.5/x)};
 \addlegendentry{$f=0.5$}
 \addplot[black,dashdotted,thick,domain=1:64,samples=100]{1/(.2+.8/x)};
 \addlegendentry{$f=0.8$}
 \addplot[black,only marks,mark=*,mark size=2pt] coordinates {(20,1.234567901234568)};
 \draw[densely dotted,gray] (axis cs:20,1) -- (axis cs:20,1.2345679);
 \node[anchor=south east,font=\scriptsize,fill=white,inner sep=1pt] at (axis cs:20,1.37) {$20\times\!\to 1.235\times$ at $f=0.2$};
\end{axis}
\end{tikzpicture}
\caption{Serial Amdahl sensitivity, not measured hardware productivity. $f$ is the baseline share of total time spent generating candidates; the marked example shows why a $20\times$ local improvement may yield only a $1.235\times$ accepted-change speedup.}
\label{fig:amdahl}
\end{figure}
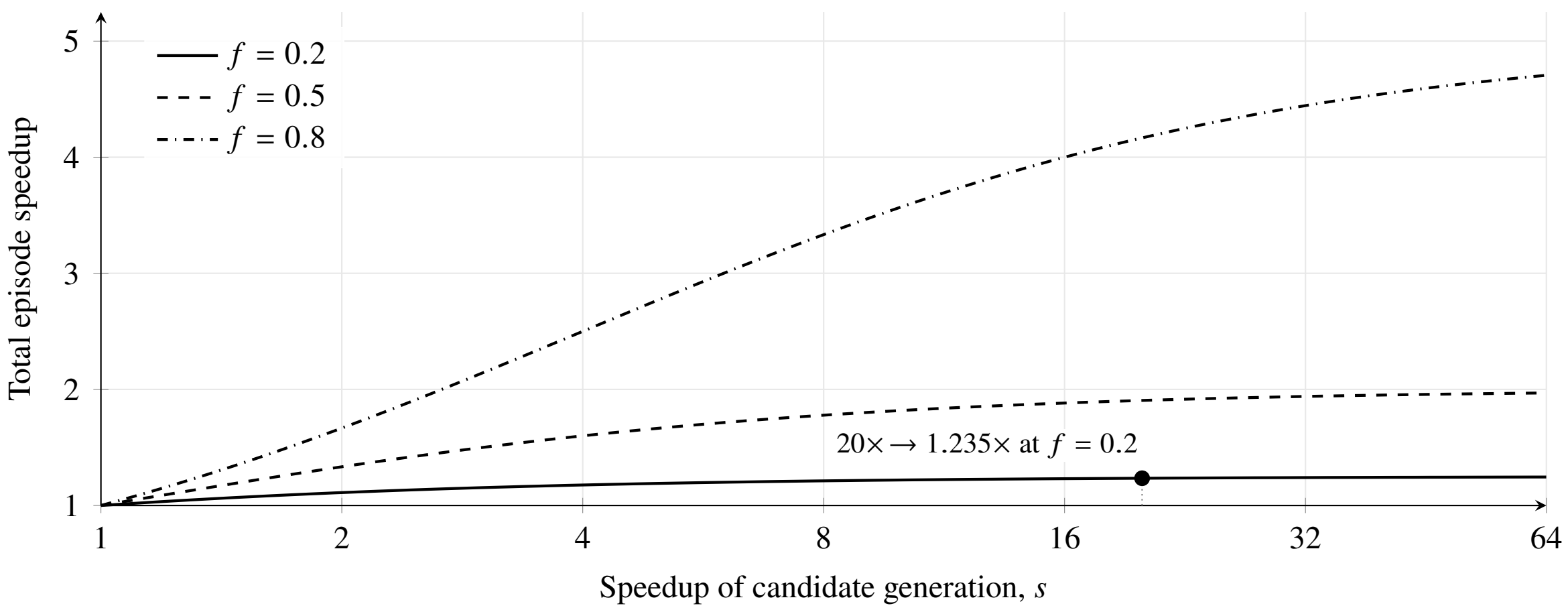

Real workflows are not serial. Their elapsed time depends on the critical path through a task graph, resource contention, and human availability. Overlapping generation and verification can reduce the critical path even when total work is unchanged. Conversely, a faster generator may have no elapsed-time effect when its output was already ready before a mandatory review or tool slot. A serious evaluation must therefore retain task timestamps and dependencies rather than infer TTAC by summing independent tool runtimes.

\subsection{An invalidation-weighted cost model}
For an elementary serial model, let $G$ be candidate-construction time, $H$ non-reused coordination and review time, and $c_j$ the cost of evidence task $j$. Define
\begin{equation}
 V=\sum_{j=1}^{n}c_j,\qquad
 \Lambda(\Delta)=\frac{\sum_{j\in I_G(\Delta)}c_j}{V}.
\label{eq:locality}
\end{equation}
Here the sum is restricted to evidence tasks, $V>0$, and $0\leq\Lambda\leq1$. Smaller $\Lambda$ means more local revalidation. The weights must be explicitly identified as runtime, cost, or human effort; these are different measures. If tasks share computation, the shared work must be represented once or attributed consistently.

With generator speedup $s$ and added evidence-management overhead $O$, the toy baseline and revised costs are
\begin{align}
 T_0 &= G+H+V,\label{eq:t0}\\
 T_1 &= \frac{G}{s}+H+O+\Lambda V.\label{eq:t1}
\end{align}
Relative to a generator-only workflow that reruns all evidence, reuse helps exactly when
\begin{equation}
 O<(1-\Lambda)V.
\label{eq:breakeven}
\end{equation}
This condition matters. A careful evidence ledger can be useful for governance even when it does not improve speed. For a tiny block, its runtime and maintenance overhead may exceed the avoided verification work. For a global constraint change, $\Lambda$ may approach one. We should expect both positive and negative cases.

\begin{table}[t]
\centering
\caption{Synthetic serial episode ($G=20$, $H=10$, $V=70$). Times and overhead $O$ are arbitrary units, not measurements.}
\label{tab:analytical}
\small
\setlength{\tabcolsep}{5pt}\renewcommand{\arraystretch}{1.15}
\begin{tabularx}{\linewidth}{Yccrrr}
\toprule
\rowcolor{gray!12}\textbf{Workflow} & $s$ & $\Lambda$ & $O$ & $T_1$ & $T_0/T_1$ \\
\midrule
Baseline & 1 & 1.00 & 0 & 100.0 & 1.00 \\
Generator only & 20 & 1.00 & 0 & 81.0 & 1.23 \\
Generator with reuse & 20 & 0.25 & 4 & 32.5 & 3.08 \\
Global edit with ledger & 20 & 1.00 & 4 & 85.0 & 1.18 \\
\bottomrule
\end{tabularx}
\end{table}

\begin{figure}[t]
\centering
\begin{tikzpicture}
\begin{axis}[
 width=.9\linewidth,height=6.1cm,
 xmin=0,xmax=1,ymin=0,ymax=1,
 xlabel={Fraction of verification effort revalidated, $\Lambda$},
 ylabel={Evidence-management overhead, $O/V$},
 xtick={0,.25,.5,.75,1},ytick={0,.25,.5,.75,1},
 tick label style={font=\footnotesize},label style={font=\footnotesize},
 axis lines=left,grid=major,grid style={gray!15},clip=false]
 \addplot[draw=none,fill=gray!18] coordinates {(0,0) (0,1) (1,0)} \closedcycle;
 \addplot[black,thick] coordinates {(0,1) (1,0)};
 \addplot[black,only marks,mark=square*,mark size=2.5pt] coordinates {(.25,.057142857)};
 \node[font=\footnotesize,align=center] at (axis cs:.26,.28) {Reuse saves\\elapsed time};
 \node[font=\footnotesize,align=center] at (axis cs:.73,.71) {Overhead exceeds\\saved verification};
 \node[anchor=south west,font=\scriptsize,fill=white,inner sep=2pt] at (axis cs:.29,.07) {Table~\ref{tab:analytical} example};
 \node[rotate=-43,font=\scriptsize,fill=white,inner sep=1pt] at (axis cs:.58,.45) {$O/V=1-\Lambda$};
\end{axis}
\end{tikzpicture}
\caption{Break-even sensitivity for the serial cost model in Eq.~\eqref{eq:breakeven}. Below the diagonal, sound reuse saves more verification work than its bookkeeping costs; above it, reuse increases elapsed time. The shaded region is analytical, not an observed deployment result.}
\label{fig:reuse_frontier}
\end{figure}
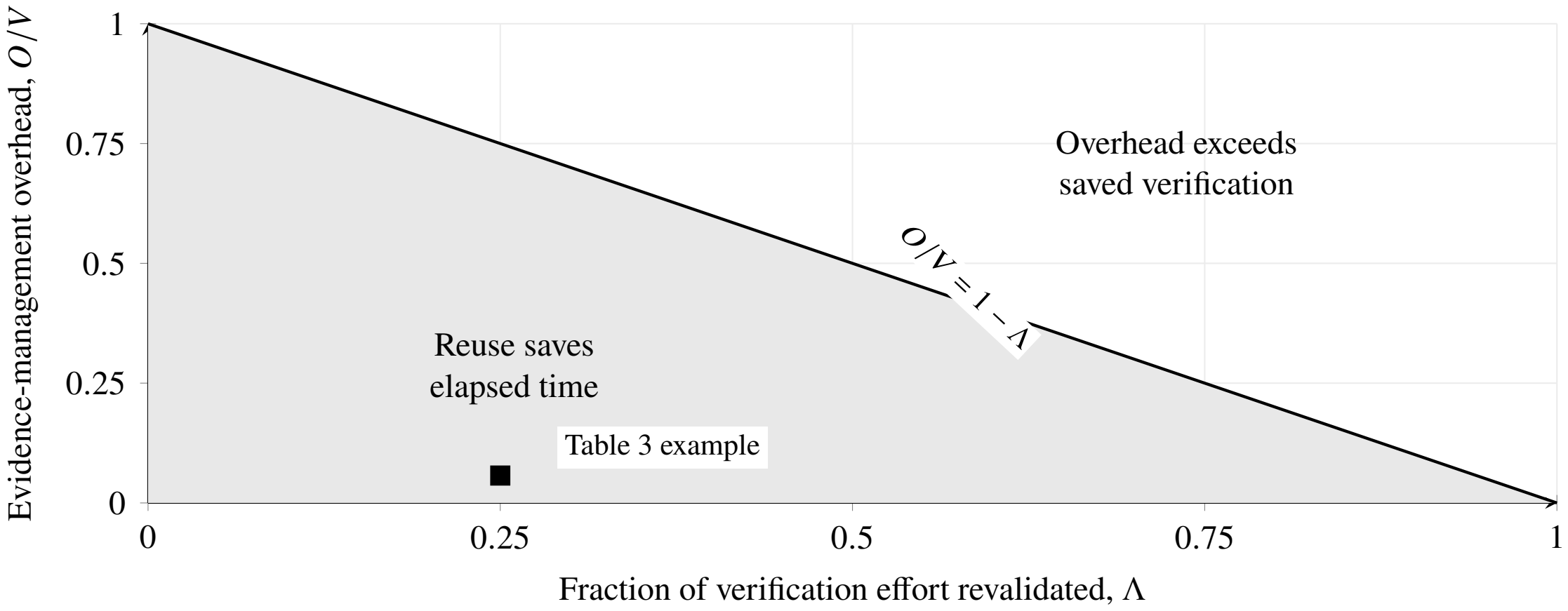

The table and \cref{fig:reuse_frontier} use arbitrary normalized time units. A reduction in $\Lambda$ is assumed, not achieved by an implemented system. It is not legitimate to advertise the resulting $3.077\times$ as an experimental speedup. Rather, the calculation identifies a measurement target: discover whether real edits permit sufficient sound evidence reuse, after accounting for setup and maintenance cost.

\subsection{The amortization horizon matters}
Let $K$ be the one-time cost of building the dependency model and protected checks, and let $b_i$ be the net benefit of reuse on change $i$. The investment breaks even only when
\begin{equation}
 \sum_{i=1}^{N} b_i > K.
\label{eq:amortization}
\end{equation}
A demonstration on a single carefully chosen edit can hide this cost. A longitudinal benchmark should include onboarding, tool-version changes, refactoring, and contract revisions, not just repeated edits inside a conveniently stable module. A reusable contract can itself become maintenance debt when requirements evolve.

\subsection{Admission control is part of agility}
Consider candidates admitted to expensive verification at rate $\lambda q$, where $\lambda$ is the generated rate and $q$ the admitted fraction. If a task consumes mean service work $\mathbb{E}[C]$ and there are $m$ equivalent workers, a basic workload condition is
\begin{equation}
 \rho=\frac{\lambda q\mathbb{E}[C]}{m}<1.
\label{eq:queue}
\end{equation}
This is a simplified capacity condition, not a complete queueing model; dependencies, heterogeneous tools, and license limits require richer analysis. Its implication is nevertheless clear: accelerating candidate production without controlling admissions can overload feedback. Cheap filters, uncertainty estimates, and staged evaluation should prioritize expensive checks. They must not silently replace a mandatory final check with a surrogate. Optimizing accepted changes per unit of scarce tool or reviewer capacity is different from maximizing proposals per hour.

\section{A worked change: the timing fix that moves the boundary}
\label{sec:example}
Consider a ready--valid streaming buffer in a single clock domain. The module accepts and emits data on rising edges when the corresponding valid and ready signals are both asserted. The producer must hold its offered data stable while stalled. A synchronous active-high reset discards in-flight transfers; conservation requirements apply separately within each interval between resets. The consumer may apply arbitrary backpressure. Eventual delivery is required only under an explicitly stated fairness assumption that the consumer continues to provide transfer opportunities.

The change request is to insert an elastic stage that removes a long combinational ready path. The contract permits additional latency within a specified bound, requires an unchanged steady-state throughput of one transfer per cycle when neither side stalls, and keeps the target clock and interface behavior fixed. This is an illustrative scenario, not a report of a real implementation experiment.

\subsection{Why an ordinary green regression is insufficient}
A candidate that merely registers ready without providing adequate buffering can accept a transfer it cannot retain when downstream readiness disappears. A short test without adversarial stalls may miss this. A reference model generated from the same mistaken interpretation may reproduce the error. Conversely, a cycle-by-cycle equality test against the old implementation may reject a correct elastic implementation because the contract intentionally permits a latency change.

The acceptance oracle must express the intended equivalence: ordered transfer traces within each reset epoch, not necessarily equality of every output on every cycle. It also needs separate obligations for stable output while stalled, legal occupancy, reset behavior, throughput, and the declared latency conditions. A conservation property alone need not detect all data corruption. A liveness claim without fairness could be false for every implementation under a permanently stalled consumer.

\subsection{How evidence should travel with the change}
The previous revision may carry a functional proof, simulation results, synthesis checks, and a post-route timing report. Editing the buffer RTL invalidates these records wherever their dependencies include the changed module. Proofs about an untouched, independently scoped decoder could remain valid, provided their parameter and environment dependencies are genuinely unchanged. A functional equivalence certificate might recover some claims, but a latency-changing transformation requires an appropriate refinement relation rather than blindly reusing a cycle-exact equivalence check.

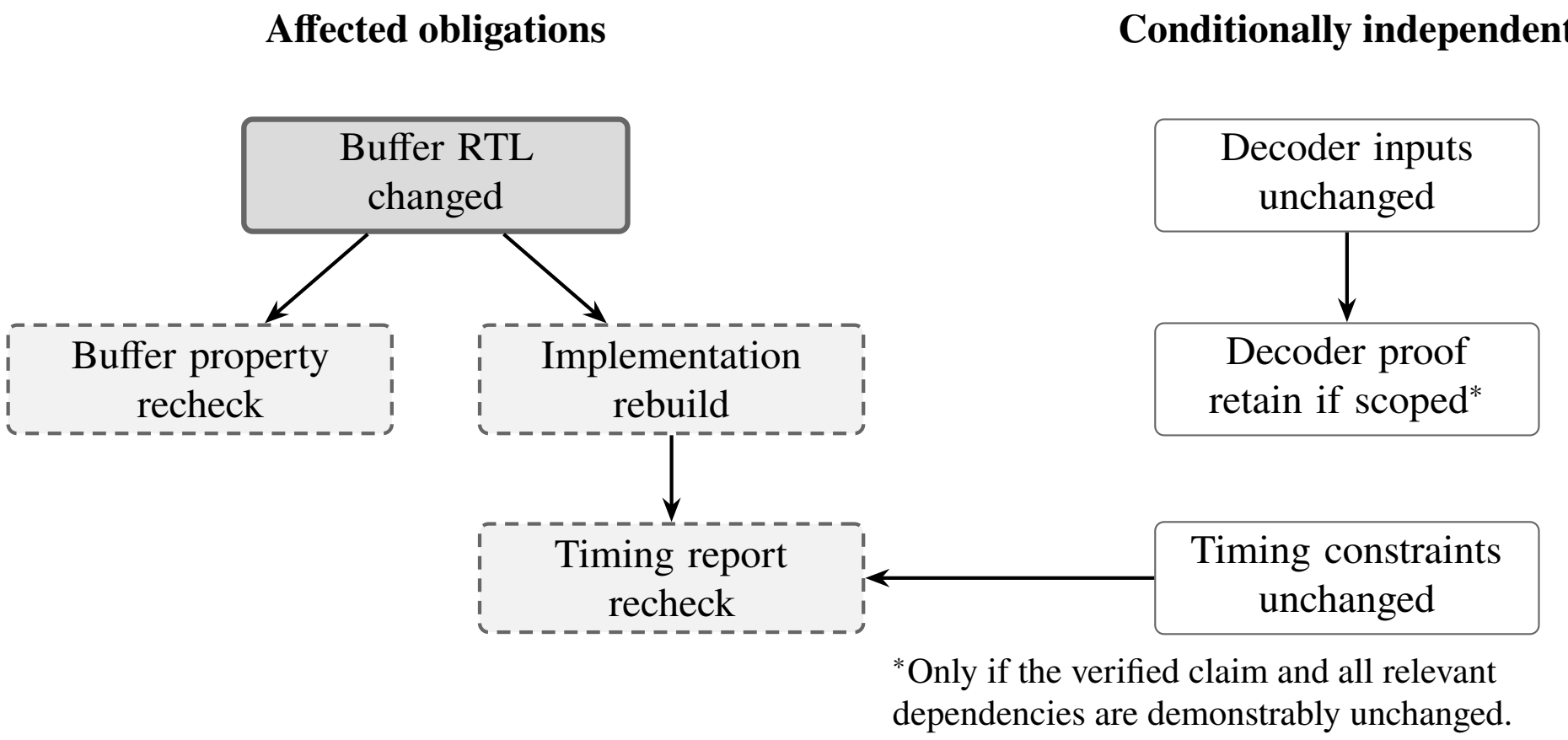
\begin{figure}[t]
\centering
\begin{tikzpicture}[
 x=1cm,y=1cm,font=\footnotesize,
 base/.style={draw=black!60,rounded corners=2pt,fill=white,align=center,text width=2.8cm,minimum height=.85cm,inner sep=3pt},
 changed/.style={base,fill=gray!28,very thick},
 rerun/.style={base,fill=gray!10,densely dashed,thick},
 keep/.style={base,fill=white},
 edge/.style={-{Stealth[length=1.9mm]},thick}]
 \node[font=\bfseries\footnotesize] at (3.2,.72) {Affected obligations};
 \node[font=\bfseries\footnotesize] at (10.35,.72) {Conditionally independent};
 \node[changed] (rtl) at (3.2,-.4) {Buffer RTL\\changed};
 \node[rerun] (property) at (1.35,-2.0) {Buffer property\\recheck};
 \node[rerun] (netlist) at (5.05,-2.0) {Implementation\\rebuild};
 \node[rerun] (sta) at (5.05,-3.56) {Timing report\\recheck};
 \node[keep] (dec) at (10.35,-.4) {Decoder inputs\\unchanged};
 \node[keep] (proof) at (10.35,-2.0) {Decoder proof\\retain if scoped$^*$};
 \node[keep] (constraint) at (10.35,-3.56) {Timing constraints\\unchanged};
 \draw[edge] (rtl)--(property);
 \draw[edge] (rtl)--(netlist);
 \draw[edge] (netlist)--(sta);
 \draw[edge] (dec)--(proof);
 \draw[edge] (constraint.west)--(sta.east);
 \node[font=\scriptsize,align=left,anchor=west] at (6.65,-4.48) {$^*$Only if the verified claim and all relevant\\dependencies are demonstrably unchanged.};
\end{tikzpicture}
\caption{Illustrative dependency-based invalidation after a buffer edit. Dashed outlines denote obligations requiring new evidence; an unchanged constraint remains an input to a fresh timing check. Retention of an unrelated proof is conditional on dependency completeness, not source-code proximity.}
\label{fig:invalidation}
\end{figure}

\begin{table}[t]
\centering
\caption{Acceptance obligations after the illustrative elastic-buffer edit. ``Recheck'' is a requirement, not a reported execution.}
\label{tab:worked}
\footnotesize
\setlength{\tabcolsep}{4pt}\renewcommand{\arraystretch}{1.18}
\begin{tabularx}{\linewidth}{P{.215\linewidth}YY}
\toprule
\rowcolor{gray!12}\textbf{Claim} & \textbf{Required evidence and effect} & \textbf{Tempting but unsafe shortcut} \\
\midrule
Ordered transfers & Recheck a data-aware trace refinement over reset epochs & Compare every cycle, or check only transfer counts \\
Stall and capacity safety & Recheck output stability and occupancy under varying backpressure & Test only always-ready traffic \\
Liveness and latency & Recheck latency bound under unchanged, explicit fairness assumptions & Assume away all downstream stalls \\
Throughput & Recheck one transfer per cycle under sustained eligible traffic & Treat a latency reduction as proof of throughput \\
Independent decoder & Retain a proof only if its full dependency scope is unchanged & Infer independence from a small textual diff \\
Physical timing & Rebuild implementation and analyze the original clock and corners & Attach an earlier post-route report to new RTL \\
FPGA checkpoint & Bind the bitstream, board, workload, and measured result & Reuse results from the previous bitstream \\
\bottomrule
\end{tabularx}
\end{table}

Physical evidence needs particular care. An old post-route report belongs to an old implementation and constraint set. The new RTL passing simulation does not refresh it. Likewise, editing the timing constraint is a contract change, not a successful design optimization. For an FPGA checkpoint, the accepted bitstream must correspond to the identified build, constraints, and board configuration. A laboratory measurement on the previous bitstream is not evidence about the new one.

As \cref{fig:invalidation} makes explicit, the locality benefit may be modest. If the timing edit changes global placement, the physical checks may all need rerunning even while several functional obligations remain reusable. That outcome is compatible with the thesis: evidence locality is a property to measure and sometimes design for, not a benefit to presume.

\subsection{What the executable artifact demonstrates}
The companion Python artifact constructs synthetic snapshots, dependency graphs, evidence records, and acceptance contracts. It checks that a stale result cannot satisfy a new snapshot, that an unknown or bounded result cannot masquerade as an unbounded proof, and that modifying a protected constraint rejects an episode under the original contract. It also distinguishes an unchanged independent obligation from one downstream of a modified implementation.

These are tests of the proposed bookkeeping semantics. The records do not originate from real EDA tools; method labels and outcomes are supplied by the model. Consequently, passing its unit tests does not demonstrate hardware correctness, secure provenance, a sound semantic dependency extractor, or improved development speed. The model is included to make assumptions executable and expose counterexamples to a simplistic ``all checks were green once'' acceptance rule.

\section{The missing technologies are not only better generators}
\label{sec:agenda}
Evidence-carrying change suggests a research agenda more specific than ``apply an LLM to each EDA stage.'' The following directions are complementary; none alone delivers autonomous silicon engineering.

\subsection{Executable contracts with explicit uncertainty}
The most valuable specification assistant may be one that identifies ambiguity rather than immediately emits RTL. It should link natural-language statements to interface events, state invariants, temporal properties, and implementation constraints; record which translations were approved; and surface contradictions between them. An assumption ledger should distinguish a real environment guarantee from an expedient restriction introduced to make a proof converge.

The technical challenge is not simply generating more assertions. It is preserving meaning across abstraction boundaries and assessing whether the obligations discriminate intended from unintended behavior. Mutation testing, deliberately violated requirements, and independent scenario construction can probe that discrimination. None proves specification completeness. A useful research target is whether an assistant reduces the effort needed to discover missing or inconsistent obligations under a fixed, independently authored evaluation set.

\subsection{Independent acceptance without expensive duplication}
Full organizational independence is often impractical for a small team. A minimal separation can still be substantive: keep golden checks and physical constraints read-only to the implementation agent, require approval for contract changes, and retain an evaluation set unavailable during search. Tool wrappers should distinguish parser failure, timeout, counterexample, vacuity warning, and successful completion.

Method diversity can add value when failures are complementary: a protocol assertion may expose a timing of events that randomized simulation rarely exercises; differential simulation may expose an incorrect formal abstraction. But diversity is a hypothesis about error coverage, not a numerical confidence multiplier. Research should measure overlapping and distinct escapes on controlled fault families and naturally occurring bugs, including errors shared between implementation and reference models.

\subsection{Incremental assurance across representations}
Incremental builds avoid recomputation when inputs are unchanged. Incremental assurance additionally asks whether a claim survives a semantic transformation. This requires interfaces between RTL, lowered representations, netlists, physical artifacts, and the properties stated about them. Candidate mechanisms include assume--guarantee contracts, compositional proofs, equivalence certificates, parameter-family reasoning, and explicitly scoped proof reuse.

The hard cases include changes to reset assumptions, black-box models, timing exceptions, and tool versions. A tool update that changes semantics or fixes a bug may invalidate an apparently identical result. A conservative dependency graph is safe but can be too coarse to help; an aggressive graph can be fast but unsound. The research objective is a measured frontier between maintenance cost, accepted reuse, and independently tested invalidation completeness, rather than the highest reuse percentage.

\subsection{Physical feedback that respects uncertainty}
Early PPA estimates are valuable for rejecting poor candidates, but their scope must be visible. Synthesis estimates, placement-based timing, and routed results do not provide the same evidence. OpenROAD's documented flow illustrates the multiplicity of implementation stages~\cite{orfs2026}; a front-end optimization cannot simply inherit conclusions from whichever stage is cheapest.

A promising agent would schedule increasing-fidelity checks according to expected decision value, retain calibration data, and escalate when a candidate is near a constraint boundary or outside the estimator's validated domain. It would still run every check required by the acceptance contract. Research should examine whether early estimates reduce wasted expensive runs without increasing escape at the protected final gate. Surrogate accuracy averaged over a dataset is insufficient if errors concentrate near feasibility boundaries.

\subsection{Tool interfaces that expose evidence, not prose alone}
An agent needs more than a terminal and a large context window. It needs stable machine-readable objects for claims, counterexamples, coverage exclusions, timing paths, assumptions, and artifact identity. Logs remain important for diagnosis, but critical acceptance semantics should not depend on unconstrained interpretation of a paragraph.

A useful interoperability layer would connect requirement identifiers to tool invocations and outputs while preserving native details. It should make unknown outcomes and unsupported translations explicit. This resembles continuous assurance infrastructure~\cite{etb2024}, but hardware adds tight coupling to configuration, technology libraries, and physical stages. A research prototype should demonstrate loss-aware adapters for a limited tool set, rather than claim a universal evidence language before confronting incompatible semantics.

\subsection{Architectures designed for cheap justified change}
Verification cost is partly an architectural choice. Explicit protocols, compositional state boundaries, well-defined reset behavior, and stable interfaces can make evidence more local. Latency-insensitive design provides an established example of using communication structure to decouple functional behavior from particular latencies~\cite{carloni2001}. It does not remove every physical constraint, but it suggests an important question: can a hardware generator optimize not only PPA, but also the future cost of justified revision?

Such an objective is necessarily multi-dimensional. Extra buffering, observability, or modular boundaries may cost area, power, or performance. The benefit may accrue only over many revisions. An evidence-carrying IP package could include configuration-qualified properties, permitted composition rules, and a reusable verification harness. The novelty would lie in demonstrating that these assets reduce later acceptance cost under realistic changes, not in renaming documentation as evidence.

\subsection{Budgeted search with backpressure}
Hardware agents should obey resource limits as hardware pipelines do: an expensive downstream stage must be able to slow proposal generation. Scheduling policies could prioritize candidates by expected feasibility, information gain, or the amount of evidence already reusable. These estimates need uncertainty and auditability; otherwise they can systematically exclude unconventional but superior designs.

The meaningful comparison is against a competent, budget-aware baseline, not an agent that naively runs every tool after every edit. The intended result is lower TTAC or higher accepted-change throughput at a specified cost and assurance level. A system that consumes ten times more licenses to finish somewhat earlier may be valuable in one setting and unacceptable in another.

\section{How to test the position rather than illustrate it}
\label{sec:evaluation}
A perspective should make itself vulnerable to evidence. The proposed framework would be weakened if a simpler mature CI pipeline delivered comparable acceptance speed and defect detection at lower total cost. It would also be weakened if practical dependency extraction were so conservative that little evidence could be reused, or so expensive to maintain that its benefit disappeared.

\begin{table}[H]
\centering
\caption{Four falsifiable hypotheses. They are prospective research claims, not findings of this manuscript.}
\label{tab:hypotheses}
\footnotesize
\setlength{\tabcolsep}{4pt}\renewcommand{\arraystretch}{1.17}
\begin{tabularx}{\linewidth}{P{.32\linewidth}YY}
\toprule
\rowcolor{gray!12}\textbf{Hypothesis} & \textbf{Required evidence} & \textbf{Disconfirming result} \\
\midrule
\textbf{H1. Bottleneck migration.} In realistic hardware changes, generator speedup yields disproportionately smaller TTAC gains as downstream validation becomes dominant. & Matched tasks and acceptance rules; generator-only intervention; critical-path traces & Near-proportional TTAC gains without increased downstream work or altered acceptance \\
\textbf{H2. Acceptance integrity.} Protected evaluation reduces oracle tampering and stale-evidence escapes. & Controlled faults, blinded holdouts, and authority audit at matched resource budgets & Equivalent escapes and cost, or extra friction without integrity benefit \\
\textbf{H3. Revalidation locality.} Audited evidence reuse reduces amortized cost on suitable change sequences. & Multiple revisions, complete setup costs, replay and dependency checks & Reuse overhead exceeds savings or invalid reuse causes escapes \\
\textbf{H4. Admission control.} Budgeted scheduling improves accepted-change throughput under tool saturation. & Fixed tool capacity; queues, resource use, acceptance outcomes; strong scheduler baseline & No throughput gain, extra delay, or useful candidates systematically excluded \\
\bottomrule
\end{tabularx}
\end{table}

\subsection{Evaluate sequences of changes}
The unit of evaluation should be a task sequence on a versioned design, not only a single specification-to-RTL exercise. Suitable families include parameterized FIFOs and streaming components, control-state machines, protocol adapters, DMA subsystems, and selected multi-clock integration tasks. The last category needs appropriate CDC models and checks; success on a single-clock buffer does not transfer automatically.

Each sequence should mix ordinary feature changes, realistic bug repairs, refactoring, parameter changes, reset changes, constraint changes, and tool or library updates. Some edits should preserve most evidence and others should force global invalidation. Include adversarial bookkeeping cases: a stale timing report, a changed proof assumption, a wrong configuration fingerprint, a modified golden test, an inconclusive tool exit, and a result produced for a different device. These test acceptance integrity, not just design ability.

Existing benchmarks are useful sources of task structure, but they do not remove the need to construct longitudinal episodes. CLOSER-Bench is especially close in its emphasis on budgeted cross-stage closure~\cite{closer2026}. The proposed extension is to measure \emph{which previously accepted claims survive subsequent changes, whether that reuse is sound, and how it affects amortized engineering cost}. This is a complementary evaluation axis, not a claim that cross-stage agent evaluation was absent.

\subsection{Use strong, controlled baselines}
We recommend four configurations. A mature human-led flow uses version control, reproducible scripts, appropriate CI, and existing incremental capabilities. An AI-assisted flow adds candidate generation and repair to that same environment. A tool-closed agent can plan and run the same available tools, without the proposed explicit evidence-transfer layer. Finally, the evidence-carrying agent adds protected contracts, typed records, invalidation, and declared preservation rules.

The comparison must not disable caching, basic provenance, or sensible CI in the earlier baselines merely to manufacture an advantage. Record which features already exist. Use ablations to distinguish the contribution of authority separation, record typing, dependency extraction, semantic reuse, and scheduling. Where feasible, compare each agent using the same underlying model and tool interfaces. Model strength and additional tools are otherwise confounding variables.

Equal wall-clock budgets are not equal resource budgets. A factorial or explicitly constrained design should account for worker count, tool licenses, CPU and GPU time, tokens, and human attention. A workflow with more human review should not present that effort as free. Initial setup and recurring maintenance should be reported separately and together. For human studies, familiarity and learning effects require counterbalancing or randomized assignment; agent repetitions alone cannot estimate human productivity.

\subsection{Protect the oracle and measure residual failures}
Requirements, golden checks, and final physical constraints should be authored or reviewed independently of the evaluated implementation search. The final evaluator should be unable to accept an edited oracle as the original one. Hidden tests, property challenges, and targeted fault injection should remain outside the search feedback loop until the endpoint is committed.

A holdout suite estimates defects discoverable by that suite; it does not reveal all remaining bugs. Mutation results likewise depend on the mutation model and the treatment of equivalent mutants. Report the overlap between failures caught by different methods, not only aggregate pass rates. A design failing a mandatory physical constraint is not accepted merely because it improves an average PPA score. Tool timeouts, parser failures, inconclusive proofs, and unavailable resources should remain visible in the denominator.

\subsection{Report a vector of outcomes}
For a fixed contract class, report the acceptance curve
\begin{equation}
 F_{\Gamma}(t)=\Pr(\TTAC_{\Gamma}\leq t)
\label{eq:acceptancecurve}
\end{equation}
alongside the final acceptance rate. For a common observation budget $B_t$, the mean of $\min(\TTAC,B_t)$ can summarize restricted waiting time if never-accepted episodes are treated as exceeding the budget. This statistic must be accompanied by acceptance rate and failure categories; it is not ``mean successful completion time.'' Reporting only the fastest successes introduces survivor bias.

Also report human active minutes, compute and tool expenditure, queueing time, preserved versus invalidated evidence costs, independent defect findings, and PPA under matched constraints. For trajectories, report accepted changes per total elapsed time and the amortization horizon including setup. A replay audit should reconstruct the exact snapshot and records underlying each acceptance decision.

Task-paired analyses can reduce heterogeneity, with uncertainty estimated at the design or repository level rather than pretending that correlated edits are independent. Quality non-inferiority margins and speed objectives should be specified before inspecting results. A speed gain with a wider residual-failure interval is a tradeoff, not automatically an advance. Where licensing and confidentiality permit, the evaluation should release configuration files, task manifests, random seeds, tool versions, rejection logs, and records of censored episodes.

\subsection{A feasible initial study}
A first empirical project need not fabricate silicon. It can restrict the contract to functional and implementation checkpoints on open digital IP, use an open simulation, formal verification, and synthesis stack, and add an FPGA checkpoint for a small subset. The study should include at least one edit where reuse helps and one where full revalidation is required. A narrow, reproducible demonstration with honest assurance limits would be more informative than a broad claim of ``autonomous chip design'' based on hand-selected successes.

The present paper does not report that study. Its executable artifact tests the acceptance abstraction and reproduces the analytical calculations. The proposed evaluation is a research protocol, not an experiment disguised by future-tense wording.

\section{The speedup frontier: from design loops to manufactured silicon}
\label{sec:speedup}
How large might the next gains be? A useful perspective should distinguish a local task speedup from the elapsed time to an accepted change, and both from delivery of a characterized physical device. It should also make its forecast assumptions inspectable. This section offers \emph{conditional scenarios for approximately 2027--2031}, viewed from late 2026, rather than an empirical forecast with calibrated confidence intervals. The purpose is to reveal leverage and constraints, not to attach a universal date or productivity multiplier to hardware engineering.

\subsection{Vendor-reported point improvements have narrow denominators}
Recent industrial announcements indicate that the frontier is moving. Cadence reports up to $10\times$ productivity improvement on selected design and verification activities, with a customer reporting roughly $10\times$ improvement in some verification work~\cite{cadence2026}. Synopsys reports up to $50\times$ faster time to validated RTL in a demonstrated agentic verification workflow~\cite{synopsys2026}. A separate Cadence announcement describes a particular RTL validation loop reduced from five weeks to under a day~\cite{cadencevirtual2026}. These are \emph{vendor-reported and task-specific} results. The chosen baseline, workflow boundary, staffing, added compute, and acceptance criteria need not be identical; none of the figures is a measured whole-chip schedule reduction. Conversely, benchmarks of hardware repair and cross-stage closure still expose difficult integration and feedback tasks~\cite{hwe2026,closer2026}. An admissible forecast must honor both observations.

\subsection{Three transparent scenarios for the next three to five years}
For a deliberately simple serial decomposition, divide the \emph{baseline elapsed time} into specification and RTL construction ($D$), verification and debug ($V$), integration and required physical checks ($P$), fabrication and assembly including production test ($M$), and post-silicon functional characterization and qualification ($Q$). Let $w_i$ be the measured baseline time fraction, $s_i$ the future speedup within that stage, and $h$ the additional time devoted to evidence governance, integration, and acceptance, normalized by the original total. Then
\begin{equation}
 R \equiv \frac{T_{\mathrm{new}}}{T_{\mathrm{base}}}
   =h+\sum_{i\in\{D,V,P,M,Q\}}\frac{w_i}{s_i},
 \qquad \sum_i w_i=1, \quad S=R^{-1}.
 \label{eq:future_pipeline}
\end{equation}
The calendar-time saving is $100(1-R)\%$, not $100(S-1)\%$. This is a sensitivity model, not a fitted law. Actual critical paths, concurrent work, license queues, human approvals, and repeated iterations must be evaluated from task graphs. In particular, speedups cannot be summed, and a stage throughput increase cannot be automatically substituted for an equal reduction in its latency.

Panel B of \cref{tab:forecast_inputs} specifies three \emph{hypothetical} baseline mixes. For reusable digital IP, the manufacturing and post-silicon shares are zero because the endpoint is the accepted design change. For an FPGA iteration, $P$ includes board-dependent checks and there is no wafer fabrication. For an ASIC project carried through to a qualified die, fabrication and assembly are kept distinct from subsequent device characterization. The weights are deliberately chosen to show how the answer depends on the acceptance boundary; they are \emph{not industry medians}. Panel A contains equally hypothetical stage multipliers, not probabilities or extrapolations of vendor peak results. In particular, manufacturing multipliers are illustrative lead-time assumptions, not equipment-throughput claims.

\begin{table}[t]
\centering
\caption{Constructed scenario inputs. Panel A contains hypothetical stage speedups; Panel B contains assumed \emph{baseline elapsed-time shares} $w_i$ and extra overhead $h$. None is an empirically fitted industry forecast.}
\label{tab:forecast_inputs}
\footnotesize
\setlength{\tabcolsep}{4pt}\renewcommand{\arraystretch}{1.15}
\begin{tabularx}{\linewidth}{Yrrrrrr}
\toprule
\rowcolor{gray!12}\textbf{A. Stage multiplier} & $s_D$ & $s_V$ & $s_P$ & $s_M$ & $s_Q$ & \\
\midrule
Conservative & 2.00 & 1.30 & 1.05 & 1.00 & 1.05 & \\
Central & 4.00 & 2.00 & 1.25 & 1.03 & 1.20 & \\
Stretch & 8.00 & 4.00 & 1.75 & 1.10 & 1.50 & \\
\midrule
\rowcolor{gray!12}\textbf{B. Baseline mix} & $w_D$ & $w_V$ & $w_P$ & $w_M$ & $w_Q$ & $h$ \\
\midrule
Reusable IP & .25 & .55 & .20 & .00 & .00 & .05 \\
Integrated FPGA & .20 & .50 & .30 & .00 & .00 & .05 \\
ASIC to qualified die & .15 & .35 & .20 & .20 & .10 & .03 \\
\bottomrule
\end{tabularx}
\end{table}

\FloatBarrier
\begin{table}[t]
\centering
\caption{Conditional time saved for each scope and scenario from Eq.~\eqref{eq:future_pipeline}; total speedup appears in parentheses. Values are synthetic calculations, not observations or calibrated forecasts.}
\label{tab:forecast_outcomes}
\small
\setlength{\tabcolsep}{5pt}\renewcommand{\arraystretch}{1.25}
\begin{tabularx}{\linewidth}{Yccc}
\toprule
\rowcolor{gray!12}\textbf{Scope} & \textbf{Conservative} & \textbf{Central} & \textbf{Stretch} \\
\midrule
Reusable IP & $21\%$ ($1.27\times$) & $45\%$ ($1.83\times$) & $67\%$ ($3.00\times$) \\
Integrated FPGA & $18\%$ ($1.22\times$) & $41\%$ ($1.69\times$) & $63\%$ ($2.69\times$) \\
ASIC to qualified die & $14\%$ ($1.16\times$) & $32\%$ ($1.47\times$) & $50\%$ ($2.00\times$) \\
\bottomrule
\end{tabularx}
\end{table}

\begin{figure}[t]
\centering
\begin{tikzpicture}[x=1cm,y=1cm,font=\footnotesize]
  \node[anchor=west,font=\scriptsize] at (3.2,4.66) {Hypothetical scenario calculations --- not observations};
  \draw[black!65,fill=white,line width=.8pt] (4.05,4.23) circle (2.5pt);
  \node[anchor=west,font=\scriptsize] at (4.23,4.23) {Conservative};
  \filldraw[fill=gray!55,draw=black!65] (7.15,4.15) rectangle (7.31,4.31);
  \node[anchor=west,font=\scriptsize] at (7.46,4.23) {Central};
  \filldraw[fill=black,draw=black] (10.02,4.23) -- (10.10,4.33) -- (10.18,4.23) -- (10.10,4.13) -- cycle;
  \node[anchor=west,font=\scriptsize] at (10.30,4.23) {Stretch};
  \foreach \tick in {0,20,40,60,70} {
    \pgfmathsetmacro{\x}{3.20+\tick*0.146}
    \draw[gray!22] (\x,.66)--(\x,3.79);
    \draw[black!70] (\x,.64)--(\x,.55);
    \node[anchor=north,font=\scriptsize] at (\x,.51) {\tick};
  }
  \node[font=\footnotesize] at (8.31,-.03) {Elapsed time saved (\% of baseline)};
  \foreach \label/\y/\a/\b/\c in {
    {Reusable IP}/3.38/21/45/67,
    {Integrated FPGA}/2.31/18/41/63,
    {ASIC to qualified die}/1.24/14/32/50
  } {
    \node[anchor=east,font=\footnotesize] at (3.00,\y) {\label};
    \draw[gray!30] (3.2,\y)--(13.42,\y);
    \pgfmathsetmacro{\xa}{3.20+\a*0.146}
    \pgfmathsetmacro{\xb}{3.20+\b*0.146}
    \pgfmathsetmacro{\xc}{3.20+\c*0.146}
    \draw[black!45,thick] (\xa,\y)--(\xc,\y);
    \draw[black!70,fill=white,line width=.8pt] (\xa,\y) circle (2.5pt);
    \filldraw[fill=gray!55,draw=black!65] (\xb-.085,\y-.085) rectangle (\xb+.085,\y+.085);
    \filldraw[fill=black,draw=black] (\xc,\y+.115)--(\xc+.095,\y)--(\xc,\y-.115)--(\xc-.095,\y)--cycle;
    \node[anchor=south,font=\scriptsize] at (\xa,\y+.14) {\a};
    \node[anchor=south,font=\scriptsize] at (\xb,\y+.14) {\b};
    \node[anchor=south,font=\scriptsize] at (\xc,\y+.14) {\c};
  }
\end{tikzpicture}
\caption{Scenario sensitivity of time saved, computed from Eq.~\eqref{eq:future_pipeline} and the assumptions in Table~\ref{tab:forecast_inputs}. Connected marks indicate the range of three constructed scenarios, \emph{not} an estimated confidence interval. The exact multipliers appear in Table~\ref{tab:forecast_outcomes}.}
\label{fig:forecast_summary}
\end{figure}
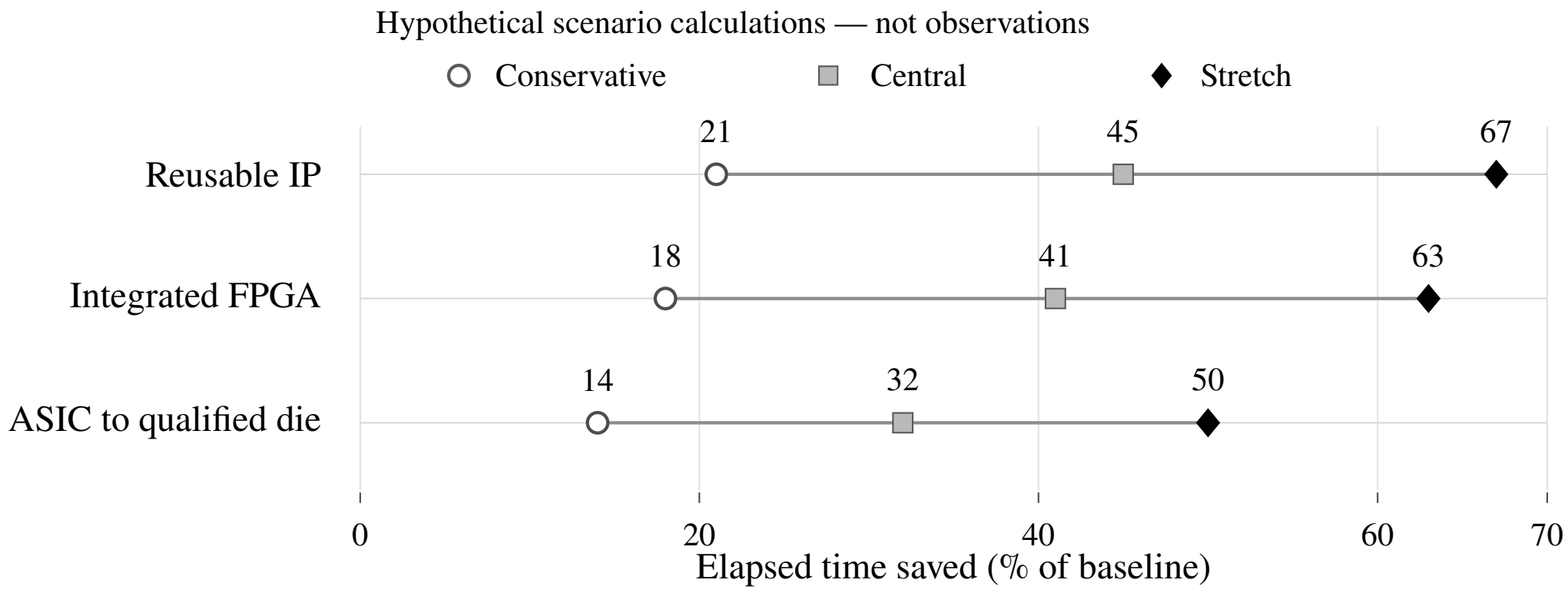

The contrasts in \cref{tab:forecast_outcomes,fig:forecast_summary} are conditional on these inputs. Under the central scenario, the illustrative accepted IP iteration is about $1.83\times$ faster, or $45\%$ shorter in elapsed time; the integrated FPGA iteration is about $1.69\times$ faster, or $41\%$ shorter; and the illustrative ASIC project to a qualified device is about $1.47\times$ faster, or $32\%$ shorter. The conservative and stretch scenarios show the sensitivity to those assumptions. These values define \emph{conditional research targets}; they are not calibrated predictions that all organizations will realize these gains by 2031. As a qualitative judgment, repeatable and well-contracted IP revisions are more promising near-term candidates for multi-fold improvement than complex ASICs that include manufacture and qualification. The stretch outcomes require favorable reuse, qualified automation, and no new unmodeled bottleneck. In particular, the constructed $s_M=1.00$--$1.10$ corresponds to a $0$--$9\%$ reduction in that stage's elapsed time, while $s_Q=1.05$--$1.50$ corresponds to roughly $5$--$33\%$ shorter post-silicon characterization. Neither interval is an empirically established sector-wide prediction.

An important stress test is an increase in design complexity or the required assurance level. If a faster generator produces more variants, stronger correctness obligations, or additional physical checks, the $w_i$ and acceptance contract do not remain fixed. A separate stress test adds evidence-ledger setup and tool-maintenance costs that recur across projects rather than changes. An attractive numerical speedup disappears if $h$ grows or if acceptance becomes more expensive. Thus, even these scenario values should be accompanied by cost and defect-escape measurements, not elapsed time alone.

\subsection{Manufacturing can accelerate, but not as software compilation does}
Physical production contains both computational preparation and serial material transformations. In computational lithography, NVIDIA, TSMC, and Synopsys reported roughly $40$--$60\times$ acceleration in specific mask-computation workflows~\cite{nvidiacuLitho2024}. That is an important example of moving a manufacturing-adjacent bottleneck, \emph{not} a corresponding acceleration of lithography equipment, wafer processing, or fab lead time. As a historical order-of-magnitude reference, the Semiconductor Industry Association described wafer fabrication as roughly 12 weeks on average, with advanced processes taking around 14--20 weeks in its 2021 account~\cite{siafab2021}. These figures are not a measurement of 2026 fab performance, but they illustrate the scale of physical and queueing processes absent from an RTL-generation benchmark.

There is still meaningful headroom in production scheduling, metrology, process control, inspection, equipment uptime, and yield learning. A 2026 industry analysis associated AI-related fab operations use cases with approximately $3$--$5\%$ higher \emph{equipment throughput} in selected applications and shorter process-change reviews~\cite{mckfab2026}. Throughput, labor productivity, mask-computation time, cycle time, and time to first customer sample are distinct denominators. For a factory with several independent constraints, a local $5\%$ throughput gain need not improve either the bottleneck tool's utilization or a particular chip's delivery date. Any claim about \emph{end-to-end manufacturing acceleration} therefore needs fab-level work-in-progress, cycle-time distributions, yield, and product-mix controls, rather than multiplying disparate stage reports.

\subsection{After fabrication, distinguish functional validation from production test}
Three commonly conflated activities deserve separate acceptance conditions. \emph{Post-silicon bring-up and functional characterization} ask whether the fabricated design works across required configurations, software workloads, voltages, temperatures, and interfaces. \emph{Production screening} at wafer sort and final test asks whether each manufactured unit passes a defect-oriented test policy. \emph{System-level test and reliability qualification} may add lengthy operating, stress, thermal, and field-representative workloads. Faster emulation, instrument orchestration, trace analysis, and test prioritization can help the first and third activities; adaptive test ordering, screening, and design-for-test can help the second. None automatically supplies evidence for the others.

Opportunities for test selection and sequencing are concrete but uneven. An NXP study described by \emph{IEEE Spectrum} considered chips undergoing 41--164 tests and proposed removing 42--74\% of test \emph{items} for some devices~\cite{mooretest2024}. It would be incorrect to convert that number directly into a $42$--$74\%$ reduction in test \emph{time}, or to assume a production-quality guarantee from the exploratory study. Research on adaptive wafer sort explicitly addresses the tradeoff among test duration, quality, and diagnostic information~\cite{adaptivewafer2025}. If screening becomes adaptive, its acceptance case must include sensitivity to drift, rare defect populations, worst-case escapes, and the preservation of diagnostic value. A cheaper test program is not an advance when it merely exports defects to customers or loses information needed for yield improvement.

\subsection{Avoided respins may dominate every local speedup}
The most consequential improvement may be a reduction in \emph{repeated physical commitments}. Write the expected delivery time schematically as
\begin{equation}
 \mathbb{E}[T_{\mathrm{delivery}}]
   = \mathbb{E}[T_{\mathrm{first\ pass}}]+\sum_{k}\Pr(E_k)\,\mathbb{E}[D_k\mid E_k],
 \label{eq:rework}
\end{equation}
where $E_k$ is a specified rework or respin event and $D_k$ is its \emph{incremental} delay; overlapping events must not be double-counted. The formula does not provide probabilities or establish that present agents reduce them. It does show why a seemingly slower, independent verification step can improve expected time to working silicon if it prevents an expensive iteration, and why test quality must not be relaxed merely to maximize apparent throughput.

\paragraph{How to test these prospective claims.}
A meaningful longitudinal study should register a baseline stage mix; preserve identical acceptance obligations across workflows; report median and tail $\TTAC$, human effort, compute and license cost; track defect escapes and respins; and identify whether benefits occur on the critical path. For silicon-facing work it should separately report time to first fabrication, fab cycle time, post-silicon acceptance, and the rate of qualified devices. Only after such measurements could the illustrative bands in \cref{tab:forecast_outcomes} be promoted to a defensible empirical forecast. The anticipated trend is therefore not that silicon manufacture becomes instantaneous, but that more invalid designs are rejected earlier, more valid evidence survives change, and fewer expensive physical iterations are needed.

\section{How far away? Capability gates before calendar claims}
\label{sec:outlook}
The conditional calculations in \cref{sec:speedup} make a numerical possibility visible, not a technology-readiness judgment. There is no single arrival date for agile hardware because the question combines different endpoints: an editable RTL candidate, a validated module, an integrated FPGA system, an ASIC implementation meeting a declared signoff contract, and reliable manufactured silicon. A model that solves the first has not necessarily progressed the last by the same fraction.

\begin{table}[H]
\centering
\caption{Observable acceptance gates across hardware scopes. These are capability boundaries, not dated forecasts.}
\label{tab:roadmap}
\footnotesize
\setlength{\tabcolsep}{5pt}\renewcommand{\arraystretch}{1.18}
\begin{tabularx}{\linewidth}{P{.225\linewidth}YY}
\toprule
\rowcolor{gray!12}\textbf{Checkpoint} & \textbf{Required capability} & \textbf{Outside the guarantee} \\
\midrule
Exploratory RTL & Reproducible build, scoped claims, protected functional checks & Integration and physical feasibility \\
Reusable digital IP & Configuration-qualified contracts, compositional assumptions, replayable evidence & Correctness for every environment or parameter \\
Integrated FPGA & Identified bitstream and board, software, timing and interface checks & ASIC equivalence or all deployment conditions \\
Digital ASIC signoff & Current functional and physical evidence at required corners and libraries & Fabrication yield or all real-world hazards \\
Manufactured system & Device and workload validation with traceable revisions & Zero escapes or fully reversible commitments \\
\bottomrule
\end{tabularx}
\end{table}

\subsection{What is already demonstrable}
The cited generation, verification, repair, and cross-stage work establishes that important parts of a design loop can be automated under specified tasks~\cite{verilogeval2023,fixme2026,hwe2026,closer2026}. The historical integrated flows establish that repeated implementation feedback is practical~\cite{lee2016,chipyard2020}. These observations justify building bounded end-to-end pilots now. They do not establish unattended reliability across arbitrary projects, nor a universal economic advantage over competent engineers.

For FPGA systems, reconfiguration makes some errors cheaper to reverse, but board interfaces, clocks, resets, firmware, deployment conditions, and observability remain part of the acceptance case. An FPGA demonstration can validate important system behaviors without serving as a complete proxy for an ASIC implementation. For ASICs, fabrication and characterization impose physical commitments that faster code generation cannot remove. Agile methods can reduce uncertainty before commitment and exploit learning after it; they cannot make every commitment reversible.

\subsection{What would count as a meaningful transition}
A credible transition to routinely agile hardware would be visible as repeatable, setup-inclusive improvements across changes and teams at a stable acceptance level. Revision histories would carry reproducible acceptance bundles. Contract modifications would be auditable. Independent evaluation would find no material degradation in the specified quality measures. Engineers would spend less time reconstructing what was tested, under which assumptions, and for which version.

The transition might arrive first in parameterized IP families, well-scoped accelerators, or FPGA subsystems whose contracts and tool environments are already explicit. That is a reasoned hypothesis about favorable conditions, not a dated market forecast. Broad autonomy over heterogeneous SoCs would require stronger evidence about compositional assumptions, integration, physical closure, and governance than small-module benchmarks can provide.

\subsection{The strongest objections}
\paragraph{``This is just good CI.''} Much of it should be. Versioning, reproducible builds, gated integration, and verification plans already exist; OpenTitan is a public example~\cite{opentitan2026,opentitandv2026}. The proposed increment is not a new label for running tests. It is explicit claim typing, freshness across semantic and physical stages, separately governed acceptance, and measurement over repeated changes. If an existing CI system already provides these capabilities at lower cost, it is a successful implementation of the principle rather than an inferior baseline.

\paragraph{``Better models will solve verification too.''} They may accelerate major parts of it. The position does not assume that verification remains a permanently fixed bottleneck. Better models can generate properties, improve abstractions, interpret counterexamples, and discover physical optimizations. The point is that their success must still be evaluated against an acceptance target they cannot quietly redefine. Faster evidence generation strengthens rather than invalidates evidence-centered measurement.

\paragraph{``The contract cannot be complete.''} Correct. An acceptance contract is a bounded engineering commitment, not omniscience. Unknown hazards, omitted requirements, model mismatch, and tool defects can remain. The framework makes the accepted scope inspectable; it does not make it exhaustive. Human judgment and independent validation remain necessary at the boundaries where formalization is incomplete or inappropriate.

\paragraph{``The overhead will overwhelm small teams.''} It may. The serial break-even condition in Eq.~\eqref{eq:breakeven} and the amortization condition in Eq.~\eqref{eq:amortization} make this objection testable. A sensible implementation starts with a few high-value obligations and stable tool adapters, then expands only when saved effort and improved integrity justify the cost. A comprehensive but rarely maintained evidence graph can be worse than a small, trustworthy one.

\paragraph{``Reuse creates a new single point of failure.''} Yes: a faulty dependency or preservation rule can systematically accept stale claims. The reuse mechanism needs its own review, adversarial tests, conservative fallback, and periodic full reruns. Full reruns are not a failure of the methodology; they are a calibration and audit instrument. A framework that cannot safely abandon reuse should not control acceptance.

\section{Conclusion: making change cheap to justify}
\label{sec:conclusion}
The history of agile hardware is a history of shortening particular loops: describing circuits, generating variants, checking properties, simulating systems, and obtaining implementation feedback. AI agents add new ways to traverse those loops, but candidate-generation speed is not a sufficient measure of progress. The appropriate unit is a change accepted under explicit, preserved conditions.

Evidence-carrying agility builds on proof-carrying artifacts, incremental computation, continuous assurance, and mature CI. Its value depends on a testable question: can sound reuse of current, scoped evidence lower the total cost of accepted changes without weakening their acceptance target? The answer depends on physical coupling, tool capacity, maintenance, and amortization. The sensitivity scenarios show why faster RTL generation, mask computation, or production-test selection cannot be converted into proportional reductions in time to qualified silicon.

The next benchmark is not merely whether an agent creates a working design, but whether it identifies which claims survive change, repairs those that do not, and supplies a reconstructible acceptance case. Generation supplies the spark. Agility starts when justified changes become quick, repeatable, and affordable.

\clearpage
\bibliographystyle{IEEEtran}
\bibliography{references}

\clearpage
\appendix
\section{A minimal evidence record and acceptance procedure}
\label{app:schema}
The following illustrative record shows the information an adapter should preserve. Identifiers are schematic, not real tool results. In production, integrity metadata would need an authenticated execution and storage mechanism; a string naming an authorized producer is not authentication.
\begin{lstlisting}
{
  "obligation": "buffer.no_lost_transfer",
  "contract_id": "approved-contract-hash",
  "artifact_id": "assessed-snapshot-hash",
  "dependencies": ["rtl", "config", "reset-model"],
  "method": "unbounded-proof",
  "tool_environment": "versioned-tool-image",
  "assumptions": "approved-environment-hash",
  "scope": "single-clock; reset-epoch semantics",
  "outcome": "PROVED",
  "witness": "content-addressed-proof-or-log",
  "producer": "authorized-verification-service"
}
\end{lstlisting}
A minimal acceptance procedure first checks protected inputs and approvals, then evaluates each mandatory obligation. A record must match the obligation's evidence type and scope, identify current dependencies, and originate from an authorized execution. Unknown or missing results remain unsatisfied. Multiple records can be supplied, but at least one admissible record is required for every obligation. Where combining records is necessary, the combination rule must itself be explicit; two insufficient records do not become sufficient by mere accumulation.

\section{Executable artifact and its limits}
\label{app:artifact}
The artifact is standard-library Python, intended for Python 3.10 or later. From the full authoring project root, or after changing into the extracted ancillary directory:
\begin{lstlisting}[language=bash]
python -m unittest discover -s artifact -p 'test_*.py' -v
python artifact/reproduce.py
python artifact/forecast_scenarios.py
\end{lstlisting}
The arXiv submission archive places the executable example under the \texttt{anc} directory, whereas the full Overleaf project contains it at the top level. The first command exercises record freshness, scope, method, protected-input, and dependency-graph rules. The second generates the analytical table and curve coordinates plus synthetic acceptance decisions in \texttt{data/}. The third recreates the conditional speedup scenarios in \texttt{data/forecast\_scenarios.csv}. No network service or EDA license is used. The LaTeX figures are generated directly from source; the manuscript does not require running Python to compile.

The dependency graph is manually declared and checked for cycles and unknown nodes. Recursive fingerprints bind synthetic source contents to derived nodes. This establishes consistency with the declared graph, not completeness of the graph. Semantic preservation certificates, real tool adapters, process isolation, cryptographic attestations, distributed storage, and human approval workflows are not implemented. The artifact must therefore not be used as a production signoff gate.

\paragraph{Artifact availability and authorship disclosure.}
The full authoring project contains the Python model, unit tests, analytical inputs, source audit, and editable figures. These source files may be supplied as ancillary material when the preprint is posted. This manuscript was drafted with generative-AI assistance; the actual authors must verify its content and references, provide their real names and affiliations, and complete funding, conflict-of-interest, and venue-specific disclosures. No authorship identity, institutional endorsement, or measured EDA improvement is implied by the illustrative artifact.

\section{A reporting record for future studies}
\label{app:reporting}
Each evaluated change should disclose its starting snapshot, request, contract version, allowed modifications, protected artifacts, and episode start boundary. Each tool event should identify the candidate, tool and environment, start and finish times, queue delay, resource use, outcome type, scope, and resulting evidence. Each acceptance should identify the exact records used and the reviewer or gate policy authorizing it.

A published summary should state the number of accepted, rejected, and censored episodes; setup and recurring effort; resource limits; hidden-evaluation findings; evidence-reuse and invalidation audit results; PPA constraints; and uncertainty intervals. Release both successful and failed trajectories where permissible. Explicitly distinguish checkpoint acceptance from tapeout approval, and tapeout approval from manufactured-device validation.

\section{Source selection and epistemic status}
\label{app:sources}
This perspective uses a selective historical and technical argument, not a systematic literature-search protocol. Sources were checked against public primary material available by November 3, 2026: research papers and author or university publication records, official project documentation, the original industry report, and vendor announcements. The source ledger in \texttt{notes/source\_audit.csv} records the role and limits of each reference. Living documentation can change after this date; repository snapshots should be pinned in an empirical follow-up.

Four categories remain distinct throughout the manuscript: reported historical or contemporary observations, deductions conditional on an explicit analytical model, proposed engineering mechanisms, and future empirical hypotheses. Vendor claims are not treated as independently replicated outcomes. The worked scenario and Python outputs belong to the analytical and proposed-mechanism categories. They supply no measured evidence about an actual language model, engineer, FPGA, or ASIC project. The added 2027--2031 scenario assumptions and their derived speedups are illustrative sensitivity inputs, not statistically calibrated forecasts of engineering productivity or manufacturing-cycle reduction. The absence of a normalized historical speed series is a limitation of the selected public evidence, not a proof that no organization possesses such data.

\end{document}